\documentclass[10pt,a4paper]{article} 
\usepackage[dvips]{graphicx}
\usepackage{amsfonts,amsmath,amssymb}
\usepackage{hyperref}

\begin{document}
\title{A Complete Spectral Analysis of the Jackiw-Rebbi Model, Including its Zero Mode}

\author{
F. Charmchi\footnote{Electronic address: f$\_$charmchi@sbu.ac.ir}~~and S. S. Gousheh\footnote{Electronic address: ss-gousheh@sbu.ac.ir}\\
 \small   Department of Physics, Shahid Beheshti University G.C., Evin, Tehran
19839, Iran}
\maketitle
\begin{abstract}
In this paper we present a complete and exact spectral analysis of  the $(1+1)$-dimensional model that Jackiw and 
Rebbi considered to show that the half-integral fermion numbers are possible due to the presence of an isolated 
self charge conjugate zero mode.
The model possesses the charge and particle conjugation symmetries.
These symmetries
mandate the reflection symmetry of the spectrum about the line $E=0$.
We obtain the bound state energies and wave functions of the fermion in this model using two different methods, 
analytically and exactly, for every arbitrary choice of the parameters of the kink, 
i.e.\;its value at spatial infinity ($\theta_0$) and its scale of variations ($\mu$).
Then, we plot the bound state energies of the fermion as a function of $\theta_0$.
This graph enables us to consider a process of building up the kink from the trivial vacuum.
We can then determine the origin and evolution of the bound state energy levels during this process.
We see that
the model has a dynamical mass generation process at the first quantized level and 
the zero-energy fermionic mode responsible for the fractional fermion number, 
is always present during the construction of the kink and its origin is very peculiar, indeed.
We also observe that, as expected, none of the energy levels crosses each other.
Moreover, we 
obtain analytically the continuum scattering wave functions of the 
fermion and then calculate the phase shifts of these wave functions.
Using the information contained in the graphs of the phase shifts and the bound states, we show that our phase shifts are consistent with      
the weak and strong forms of the Levinson theorem .
Finally, using the weak form of the Levinson theorem, we confirm that the number of the zero-energy fermionic modes is exactly one.
\end{abstract}

\section{Introduction}
\par
It is known that the modification of the spectrum of the Fermi field due to its coupling 
to other field configurations influences many properties of the system and causes
many interesting phenomena.
One of these phenomena is the appearance of nonzero and even noninteger fermion number of the vacuum.
The occurrence of the fractional fermion number was first pointed out by Jackiw and Rebbi in $1976$ \cite{jackiw-rebbi}.
They considered some models possessing the charge conjugation symmetry, 
in which the fermion is coupled to a scalar background field in the form of a soliton.
They concluded that the existence of an isolated nondegenerate zero-energy fermionic mode implies 
that the soliton is a degenerate doublet carrying fermion number $\pm \frac{1}{2}$.
Their surprising result has motivated much of the works on this subject.
This effect has been studied extensively in the literature for different physical models in 
many branches of physics such as particle physics \cite{jackiw-rebbi,jackiw,goldstone,mackenzie1,mackenzie2,dr,leila,dehghan}, 
cosmology \cite{cosm1,cosm2,cosm3,cosm4,cosm5}, condensed-matter physics \cite{cond1,cond2,cond3,cond4}, 
polymer physics \cite{poly1,poly2,poly3} and atomic physics \cite{atom1,atom2,atom3}.
\par
In the early 80s two systematic and elegant methods were developed as a way to
evaluate the vacuum polarization of the fermions induced by the presence of prescribed static background fields.
The first method called the adiabatic method, invented by Goldstone and Wilczek \cite{goldstone}, basically 
consists of building up adiabatically the final configuration of the background field starting from the free vacuum.
Then, the charge of the final state can be obtained by observing the fermionic current at spatial infinity, computed 
from the lowest order Feynman loop diagram.
This method is limited to the slowly varying background fields.
For infinitely slow variations of the background, none of the bound states of the fermion crosses the line of $E=0$.
Consequently, there exists only one contribution to the vacuum polarization, i.e.\;the adiabatic contribution.
This contribution can be attributed to the change in the number of continuum states with negative energy 
and is responsible for the fractional part of the fermion number.
\par
The second method was invented by MacKenzie and Wilczek \cite{mackenzie1,mackenzie2}, and the restriction of 
adiabaticity is lifted.
To calculate the charge of the no particle state in the presence of an interaction, 
one starts with the definition of the particle number operator in the free Dirac case and transforms it into the one 
which is appropriate for the basis of states in the presence of the interaction.
Then, the vacuum charge turns out to be equal to the difference between the number of negative energy states in the 
presence and absence of the background field.
One can then return back to the first quantized level and solve for 
the spectrum of the Dirac field in the presence of the background field. 
As we couple the Dirac field 
to an external potential, the Dirac equation is altered and 
the spectrum of the fermion is distorted.
The positive and negative continua change and bound states may appear.
When the scale of spatial variation of the background field is much larger than the Compton wavelength of the fermion 
($\lambda$), i.e.\;the adiabatic regime, the results of the two methods coincide.
However, when the aforementioned scale becomes comparable to or smaller than $\lambda$, the bound state energy levels could 
cross the $E=0$ line, and the definition of the vacuum changes.
This contribution to the vacuum polarization is called the nonadiabatic contribution.
In this sense the second method generalizes the first.
\par
In this paper we concentrate on
the simple but important model considered by Jackiw and Rebbi.
In this model a Fermi field is coupled to an external scalar field in the form of kink in 
$(1+1)$ dimensions.
This system has charge conjugation and particle conjugation symmetries.
These symmetries 
relate each state of the fermion with positive energy 
$E$ to a state with the energy $-E$.
Therefore, the whole spectrum of the system is completely symmetric with respect to 
the line of $E=0$.
Jackiw and Rebbi stated that there is an isolated zero-energy fermionic mode in this
system, which is self charge conjugate and showed that the vacuum polarization of the
system due to the presence of this mode is $\pm\frac{1}{2}$.
They have obtained the exact form of the zero mode \cite{jackiw-rebbi,jackiw}.
Before these works, the same model had been discussed by  Dashen {\it et al.} in $1974$ \cite{dashen} 
in a different context and with a different purpose.
In that paper, they extended the semiclassical method  to include fermions.
They also pointed out the existence of the zero-energy fermionic mode and obtained an expression for the 
discrete bound levels.
Later on Rajaraman gave the same expression for the bound state energies of the Jackiw-Rebbi model while reviewing the previous works \cite{rajaraman}.
However, as far as we know, exact solution for the whole spectrum of this model, which was introduced 
four decades ago has been heretofore missing.
In this paper we calculate the exact spectrum of the system, i.e.\;the eigenstates in the continua and all of the 
bound eigenstates and their energies, for the whole allowed ranges of the parameters, i.e.\;$\theta_0$ which 
denotes the value of the kink at $x\to\infty$ and $\mu$ the slope at $x=0$.
Having such solutions, we can explore what exactly happens to the spectrum of the fermion as the background 
field evolves from the trivial vacuum to the kink. 
From this evolution perspective, the origin of an isolated zero mode in a system with 
particle conjugation symmetry has been a  mystery to us.
The symmetries of the system clearly mandate the zero mode to be self charge conjugate, but disallow 
any levels crossing $E=0$ during the evolution process, and this obfuscates the mystery even further.
When this analysis illuminates the origin of this zero mode, some of its unexpected features become manifest. 
To this end, we obtain the bound states of the fermion for this system using two different analytical methods.
In the first method we solve the equations of motion, directly.
Our second method is the elegant shape invariance method \cite{shape1,shape2,shape3,shape4,shape5,shape6,shape7}.
Shape invariance indicates the presence of an integrability condition for the potential of the Schr\"{o}dinger-like equations.
When such a condition is satisfied for the potential, one can use the supersymmetry algebra 
to exactly solve the Schr\"{o}dinger-like equation and obtain the stationary states and their corresponding energy eigenvalues.
Our second order equations, obtained by decoupling the two first order equations embedded in the Dirac equation, 
turn out to be two Schr\"{o}dinger-like equations which are surprisingly partner Hamiltonians and related to each other through 
a shape invariance condition.
Therefore, we are able to easily obtain the solutions for our bound states using the shape invariance method.
We plot the bound state energies of the fermion as a function of $\theta_0$.
The invariance of the system under the charge and particle conjugation symmetries is obvious in this graph.
We also obtain the continuum scattering states of the fermion by solving the 
equations of motion directly, and then calculate the phase shift of these states.
One of our observations is that there is a dynamical mass 
generation process operating at the first quantized level. 
We check the consistency of our results with both the weak and strong forms of the Levinson theorem 
\cite{levinson,levinsondr}, and use those two forms to gain further insight into the fermionic spectrum.
\par
The outline of the paper is as follows.
In section $2$ we introduce the model.
In section $3$ we obtain the bound states of the fermion using two different methods.
In section $4$ we compute the continuum wave functions of the fermion and then calculate the phase shift of these states.
Then, we check the consistency of our results with both the weak and strong forms of the Levinson theorem.
In section 5 we summarize and discuss the results and draw some conclusions.
\section{The preliminaries of the Jackiw-Rebbi model}
In this section we review the basic definitions of the $(1+1)$-dimensional model studied by Jackiw and Rebbi \cite{jackiw-rebbi}.
This model includes a spinor field $\psi$ coupled to a scalar field $\phi$ through the following Lagrangian
\begin{equation}\label{lagrangian}\vspace{.2cm}
 {\cal L}=\bar{\psi}\left[i\gamma^{\mu}\partial_{\mu}-
g \phi_{\mathrm{cl}}(x)\right]\psi,
\end{equation}
where $g>0$ and $\phi_{\mathrm{cl}}(x)$ is a prescribed pseudoscalar field and chosen to be 
the kink of the $\phi^4$ theory, 
i.e.\;$\phi_{\mathrm{cl}}(x)=\left(m/\sqrt{\lambda}\right)\tanh\left(m x/\sqrt{2}\right)$.
Two important parameters which describe the kink are $\theta_0=\phi_{\mathrm{cl}}(\infty)=\frac{m}{\sqrt{\lambda}}$ 
and $\mu=\frac{\textrm{d}\phi_{\mathrm{cl}}}{\textrm{d}x}\big|_{x=0}=\frac{m^2}{\sqrt{2\lambda}}$.
Notice that the Lagrangian has no explicit fermion mass term and the mass of the free fermion is obviously zero.
However, as we shall show, the interaction term of this Lagrangian gives the mass 
$M_\mathrm{f}=|g\langle \phi_{\mathrm{cl}}\rangle|=g\frac{m}{\sqrt{\lambda}}=g\theta_0$ to the fermion, to lowest order. 
\par
We use the representation $\gamma^0=\sigma_1$ and $\gamma^1=i\sigma_3$ for
the Dirac matrices and represent the Fermi field by $\psi(x,t)=\mathrm{e}^{-i Et}\left(\! \begin{array}{c}
\psi^{(+)}(x)\\
\psi^{(-)}(x)
\end{array}\! \right)$.
Then, the Dirac equation in the presence of the background field
$\phi_{\mathrm{cl}}(x)$ can be written as follows
\begin{equation}\label{eom}
\left(
\begin{matrix}
-\partial_x-g\phi_{\textmd{cl}}(x) & E \\
E & \partial_x-g\phi_{\textmd{cl}}(x)
\end{matrix}
\right)
\left(
\begin{matrix}
\psi^{(+)}(x) \\
\psi^{(-)}(x)
\end{matrix}
\right)=0.
\end{equation}
Our purpose is to solve this equation exactly and explore the results in detail.
Before doing that, we state some important symmetries of this model.
It possesses the charge conjugation symmetry.
In our representation the charge conjugation operator includes $\sigma_3$ and
it relates the states with positive energy to the ones with negative energy as
$\psi^c_{-E}=\sigma_3\psi^*_E$.
Also, if there is a zero-energy fermionic mode, it is self conjugate, 
i.e.\;$\psi^c_{0}=\sigma_3\psi^*_0=\psi_0$.
This model also possesses particle conjugation symmetry.
In our representation the particle conjugation operation is given by $\psi_{-E}=\sigma_3\psi_E$.
Therefore, for every state with positive energy $E$ there exists a corresponding state with energy $-E$.
The chosen model is not invariant under the parity, since the background field $\phi_{\mathrm{cl}}(x)$ is the kink which is an odd function in space.
Therefore, this model does not preserve the CP and consequently it is not invariant under the time reversal.
\section{Bound states of the fermion in the presence of the background field}
In this section we solve Eq.\,(\ref{eom}) to find the wave functions of the bound states along with their associated discrete energies.
First, we solve the equations of motion directly.
Then, we solve the equations using the formalism of shape invariance, as a double check.
For an alternative derivation of the bound states see \cite{arxiv}.
\subsection{The direct method}
Equation (\ref{eom}) consists of two coupled first order differential equations. 
We first solve the two decoupled second order equations and then look for a set of solutions which are consistent 
with the original Dirac Eq.\,(\ref{eom}).
The decoupled equations are as follows
\begin{align}\label{new2ndorder}
 \frac{\textmd{d}^2\psi^{(\pm)}(x')}{\textmd{d} x'^2}+\left[\epsilon_{\pm}
-v_{\pm}~\textmd{tanh}^2\left(x'\right)\right]\psi^{(\pm)}(x')=0,
\end{align}
where we have rescaled the original parameters of the model as follows:
$x'=(\mu/\theta_0)x$, $g'=(\theta_0^2/\mu)g$ and $E'=(\theta_0/\mu)E$.
We have also defined two new parameters as follows:
$\epsilon_{\pm}=\left(E'^{\pm}\right)^2\pm g'$ and
$v_{\pm}=g'^2\pm g'$.
The solutions to these two Schr\"{o}dinger-like equations can be inferred from some old literature \cite{morse,landau}.
However, we present a very short derivation of the solutions, which we shall later combine to obtain the solution to the original Dirac equation.
When $\epsilon_{\pm}<v_{\pm}$, these equations have solutions vanishing at spatial infinities and 
consequently for this range of parameters we can have bound states. However, for $\epsilon_{\pm}>v_{\pm}$ 
the continuum solutions which are oscillatory at spatial infinities are possible.
To solve these equations, we substitute the ansatz
$\psi^{(\pm)}(x')=\textmd{sech}^{b_{\pm}}\left(x'\right)F_{\pm}(x')$ into
Eq.\,(\ref{new2ndorder}) and obtain
\begin{align}\label{subs}
&\textmd{sech}^{b_{\pm}}\left(x'\right)
\bigg\{\frac{\textmd{d}^2F_{\pm}(x')}{\textmd{d}x'^2}
-2 b_{\pm}~\textmd{tanh}\left(x'\right)\frac{\textmd{d}F_{\pm}(x')}{\textmd{d}x'}\nonumber\\
&+\left[\epsilon_{\pm}-v_{\pm}+b_{\pm}^2
+\left(v_{\pm}-b_{\pm}(b_{\pm}+1)\right)\textmd{sech}^2\left(x'\right)\right]F_{\pm}(x')\bigg \}=0.
\end{align}
In this equation the terms in the curly bracket should add up to zero.
By choosing the arbitrary parameters $b_{\pm}$ such that $\epsilon_{\pm}-v_{\pm}+b_{\pm}^2=0$, and changing of variable
$u=\frac{1}{2}\left[1-\textmd{tanh}\left(x'\right)\right]$,
the differential equations for $F_{\pm}(x')$ turn into a hypergeometric equation
with the following general solution
\begin{align}\label{oursol}
&A~{}_{2}F_{1}\left(b_{\pm}+\frac{1}{2}-\sqrt{v_{\pm}+\frac{1}{4}}
,b_{\pm}+\frac{1}{2}+\sqrt{v_{\pm}+\frac{1}{4}},1+b_{\pm}~;u\right)\nonumber\\
&+B~u^{-b_{\pm}}{}_{2}F_{1}\left(\frac{1}{2}-\sqrt{v_{\pm}+\frac{1}{4}}
,\frac{1}{2}+\sqrt{v_{\pm}+\frac{1}{4}},1-b_{\pm}~;u\right).
\end{align}
For the bound states we set $b_{\pm}>0$ to turn $\textmd{sech}^{b_{\pm}}\left(x'\right)$ into a damping factor. 
However, since $b_{\pm}>0$, $\lim_{u \to 0}u^{-b_{\pm}}=\infty$ and 
we have to set $B=0$.
In order that the remaining solution have the proper asymptotic behavior, we have to impose the 
following constraint
\begin{align}\label{constraint}
b_{\pm}+\frac{1}{2}-\sqrt{v_{\pm}+\frac{1}{4}}=-n,
\end{align}
where $n$ is a semi-positive integer. 
This constraint along with the constraint 
$\epsilon_{\pm}-v_{\pm}+b_{\pm}^2=0$ determine the
allowed discrete energies of the system. 
Using these two constraints, and the definitions of $v_{\pm}$ and 
$\epsilon_{\pm}$, the allowed energies
in terms of the original parameters $\theta_0$ and $\mu$ are as follows
\begin{align}\label{bnd}
E_{n}^{+}&=\pm\sqrt{2g\mu n-\frac{\mu^2}{\theta_0^2}n^2}
,~~~~n=0,1,2,\dots<\frac{g\theta_0^2}{\mu},\\\label{bnd2}
E_{n}^{-}&=\pm\sqrt{2g\mu n-\frac{\mu^2}{\theta_0^2}n^2},
~~~~n=1,2,\dots<\frac{g\theta_0^2}{\mu}.
\end{align}
We have reverted to the original parameterization since it is necessary for explaining
the origin of the zero mode.
The upper bounds for the integer $n$ have been obtained using the
constraint $b_{\pm}>0$.
The corresponding wave functions are
\begin{align}\label{waves}
\psi_{n}^{(+)}(x)& =N_{+}\left[\textrm{sech}\left(\frac{\mu}{\theta_0}x\right)\right]^{\frac{g\theta_0^2}{\mu}-n}
     \,{}_{2}F_{1}\left(-n,2\frac{g\theta_0^2}{\mu}-n+1,\frac{g\theta_0^2}{\mu}-n+1
     ;\frac{1}{1+\textmd{e}^{\frac{2\mu}{\theta_0}x}}\right),\\
\psi_{n}^{(-)}(x)&
  =N_{-}\left[\textrm{sech}\left(\frac{\mu}{\theta_0}x\right)\right]^{\frac{g\theta_0^2}{\mu}-n}
\,{}_{2}F_{1}\left(-n+1,2\frac{g\theta_0^2}{\mu}-n,\frac{g\theta_0^2}{\mu}-n+1
  ;\frac{1}{1+\textmd{e}^{\frac{2\mu}{\theta_0}x}}\right).\label{waves2}
\end{align}
\par
Now, we use these solutions to construct the solutions of the original Dirac's coupled first order differential equations (\ref{eom}). 
It is important to note that the energy $E$ of both of these solutions for any state should be the same, 
since $E$ is the energy of the fermion and its wave function is the doublet 
$\mathrm{e}^{-i Et}\left(\! 
\begin{array}{c}
\psi^{(+)}(x)\\
\psi^{(-)}(x)
\end{array}\! \right)$.
We define $E_{n}=E_{n}^{+}=E_{n}^{-}$ and choose the wave function of the fermion to be in the
following form
\begin{align}\label{doublet}
\psi_n(x,t)=\textmd{e}^{-iE_{n}t}\left(
\begin{matrix}
\psi_{n}^{(+)}(x) \vspace{.2cm}\\
\psi_{n}^{(-)}(x)
\end{matrix}
\right),~~~~n=0,1,2,\dots<\frac{g\theta_0^2}{\mu}.
\end{align}
One can easily check that the solutions given by Eq.\,(\ref{doublet}) satisfy 
the first order Eq.\,(\ref{eom}), if we set $N_-/N_+=n \mu/(\theta_0 E_{n})$.
Therefore, the total number of the bound states for a given value of $\theta_0$ is 
$N_{\mathrm{b}}=2\left[\frac{g \theta_0^2}{\mu}\right]_{\mathrm{F}}+1$, where $[f(\theta_0)]_{\mathrm{F}}$ denotes the floor function.
\par
Now let us concentrate on the zero-energy mode since this is the most important bound state. 
For the $E=0$ mode, $n=0$ and only the upper component is nonzero. 
We can easily extract the explicit form of the zero mode from our exact solution given in Eqs.\,(\ref{bnd}-\ref{doublet}).
Due to the importance of the zero mode, let us find it by a second method which we choose to be 
a direct calculation based on the Dirac Eq.\,(\ref{eom}) (see \cite{rajaraman}).
Setting $E=0$ in the first order Eq.\,(\ref{eom}), the two equations decouple and 
their solutions are as follows
\begin{align}\label{zero}
\psi_0^{(+)}(x)=c_{+} \left[\cosh\left(\frac{\mu}{\theta_0}x\right)\right]^{-{\frac{g\theta_0^2}{\mu}}},~~~~
\psi_0^{(-)}(x)=c_{-} \left[\cosh\left(\frac{\mu}{\theta_0}x\right)\right]^{{\frac{g\theta_0^2}{\mu}}},
\end{align}
where $c_{+}$ and $c_{-}$ are constant.
Since $\psi_0^{(-)}(x)$ makes the fermion wave function for the zero-energy mode unnormalizable, we set $c_{-}=0$.
Therefore, the wave function for this mode is
\begin{equation}\label{zeromode}
\psi_0(x)=c_{+}
\left(
\begin{matrix}
\left[\cosh\left(\frac{\mu}{\theta_0}x\right)\right]^{-{\frac{g\theta_0^2}{\mu}}}\vspace{0.2cm}\\
0
\end{matrix}
\right).
\end{equation}
We should mention that the asymptotic behavior of all of the bound states can be easily obtained from Eq.\,(\ref{new2ndorder}) and are as follows
\begin{align}
\lim_{|x|\to\infty}\psi_n^{(\pm)}(x)= \mathrm{e}^{-b_n^{\pm}\frac{\mu}{\theta_0}|x|}.
\end{align}
This asymptotic behavior exactly matches the corresponding behavior of our exact solutions given in Eqs.\,(\ref{bnd}-\ref{doublet}).
\subsection{The shape invariance method}
In this subsection we use the shape invariance method to re-derive the bound state sector of this model.
Since there are many excellent reviews on the shape invariance method (see for example \cite{shape3}), 
we shall only use the results, with minimal introduction.
Consider a sequence of Hamiltonians which are related to each other by the shape invariance condition,
\begin{align}\label{shapeinvariance2}
H_{k}(g_{1})=H_1(g_k)+\sum_{s=1}^{k-1} R(g_s),~~~~k=2,3,\dots,p,
\end{align}
where $p$ is the number of bound states of $H_1$, $R(g_s)$ is a c-function depending on the 
parameters of the Hamiltonians, and  $g_s=f^{s-1}(g_1)$.
The Hamiltonian $H_1$ and its ground state are required to have the following properties
$H_1(g)=A^{\dag}(g)A(g)$, and $A(g_1)\psi_{0}^{(1)}(g_1)=0$. 
The $n$th  Hamiltonian $H_n$ in this sequence has the same spectrum as $H_1$ except that the first $n-1$ 
bound states of $H_1$ are absent in the spectrum of $H_n$.
Using this relation, it is obvious that the ground state of the Hamiltonian $H_k$ is the $k$th energy level of $H_1$.
The bound state energies of $H_1$ are as follows
\begin{align}\label{spectrumh1}
E_{n}^{(1)}=\sum_{s=1}^{n}R(g_s),~~~~n=1,2,\dots,p,~~~\mathrm{and}~~~E_{0}^{(1)}=0.
\end{align}
Moreover, the $n$th eigenstate of $H_1$ is as follows
\begin{align}\label{waveh1}
\psi_{n}^{(1)}(g_1)\propto A^{\dag}(g_1)A^{\dag}(g_2)\dots A^{\dag}(g_n)\psi_{0}^{(1)}(g_{n+1}),~~~~n=1,2,\dots,p.
\end{align}
\par
Now, we use the shape invariance method to solve the second order
 Schr\"{o}dinger-like equations (\ref{new2ndorder}). 
We can write the partner Hamiltonians for our equations, in terms of the parameters $\mu$ and $\theta_0$, as follows
\begin{align}\label{partnerh3}
H_+&=A^{\dag}A=-\frac{\textmd{d}^2}{\textmd{d} x^2}+g^2\theta_0^2 -g \theta_0\left(g \theta_0+\frac{\mu}{\theta_0}\right)\mathrm{sech}^2\left(\frac{\mu}{\theta_0}x\right)
,\\\label{partnerh4}
H_-&=AA^{\dag}=-\frac{\textmd{d}^2}{\textmd{d} x^2}+g^2\theta_0^2 -g \theta_0\left(g \theta_0-\frac{\mu}{\theta_0}\right)\mathrm{sech}^2\left(\frac{\mu}{\theta_0}x\right),
\end{align}
where, the lowering and raising operators which build $H_+$ and $H_-$ are as follows
\begin{align}\label{aadag}
A&=-\frac{\textmd{d}}{\textmd{d} x}-g \theta_0\,\textmd{tanh}\left(\frac{\mu}{\theta_0}x\right)
,\\\label{aadag2}
A^{\dag}&=\frac{\textmd{d}}{\textmd{d} x}-g \theta_0\,\textmd{tanh}\left(\frac{\mu}{\theta_0}x\right).
\end{align}
It can be easily seen that the Hamiltonians (\ref{partnerh3}) and (\ref{partnerh4}) satisfy the following shape invariance condition
\begin{align}\label{relation}
H_-(g)=H_{+}\left(g-\frac{\mu}{\theta_0^2}\right)+\left(2 g \mu-\frac{\mu^2}{\theta_0^2}\right).
\end{align}
Thus, in our case $f(g)=g-\frac{\mu}{\theta_0^2}$ and $R(g)=2 g \mu-\frac{\mu^2}{\theta_0^2}$.
Using these two functions and Eq.\,(\ref{spectrumh1}), the energies of  the bound states of $H_+$ are as follows
\begin{align}\label{spectrum2}
E_n^{+}&=\pm\sqrt{2 n g \mu-n^2\frac{\mu^2}{\theta_0^2}}\,,~~~~n=0,1,2,\dots.
\end{align}
Substituting Eqs.\,(\ref{zeromode}) and (\ref{aadag2}) into Eq.\,(\ref{waveh1}), 
the bound state wave functions of the Hamiltonian $H_+$ can be easily obtained.
The ground state $\psi_0^{(+)}$ is already shown in Eq.\,(\ref{zeromode}) (upper component) 
and the first two excited states are as follows
\begin{align}\label{spectrum}
\psi_1^{(+)}(x)&\propto\sinh\left(\frac{\mu}{\theta_0}x \right) \left[\cosh\left(\frac{\mu}{\theta_0}x \right)\right]^{-\frac{g\theta_0^2}{\mu}},\nonumber\\
\psi_2^{(+)}(x)&\propto
\left[g\theta_0^2+\left(\mu-g\theta_0^2\right)\cosh\left(\frac{2\mu}{\theta_0}x\right)\right]
 \left[\cosh\left(\frac{\mu}{\theta_0}x \right)\right]^{-\frac{g\theta_0^2}{\mu}}.
\end{align}
As it can be seen, these energies and states are exactly the ones we obtained in the previous subsection.

\section{Continuum scattering states of the fermion in the presence of the background field} 
Now, we focus our attention on obtaining the continuum scattering states for this model.
As in the case of bound states, we choose the continuum wave functions to be in the form of $\psi_k^{(\pm)}(x)=\textmd{sech}^{b_{\pm}}\left({\mu
x}/{\theta_0}\right)F_{\pm}(x)$.
From the constraint $\epsilon_{\pm}-v_{\pm}+\frac{\mu^2}{\theta_0^2}b_{\pm}^2=0$, we have $b_{\pm}=\sqrt{\frac{\theta_0^2}{\mu^2}(v_{\pm}-\epsilon_{\pm})}$.
To have oscillatory behavior in spatial infinities for the continuum states, we set $b_{\pm}=-ik$ in which $k$ is a real quantity. 
This choice corresponds to the region $\epsilon_{\pm}>v_{\pm}$.
Using the definitions of the parameters $v_{\pm}$ and $\epsilon_{\pm}$ in terms of $\theta_0$ and $\mu$, we obtain $E^2=\frac{\mu^2}{\theta_0^2}k^2+g^2\theta_0^2$.
This equation reveals an extremely important property of this model, 
namely dynamical mass generation at the tree level.
This mass is given by $M_\textrm{f}=g\theta_0$.
We shall comment further on the properties of the spectrum of this model in subsection $4.1$.
\par
The continuum states of the fermion are as follows
\begin{align}\label{scatt}
\psi^{(\pm)}_{k,\mathrm{L}}(x)
=N_{\pm}^{k,\mathrm{L}}\textrm{cosh}^{ik}\left(\frac{\mu}{\theta_0}x\right)
{}_{2}F_1\left(\frac{1}{2}-ik-\zeta_{\pm} ,\frac{1}{2}-ik+\zeta_{\pm} ,1-ik
;\frac{1}{1+\textrm{e}^{\frac{2\mu}{\theta_0}x}}\right),
\end{align}
\begin{align}\label{scatt2}
\psi^{(\pm)}_{k,\mathrm{R}}(x)=N_{\pm}^{k,\mathrm{R}}\textrm{cosh}^{ik}\left(\frac{\mu}{\theta_0}x\right)
{}_{2}F_1\left(\frac{1}{2}-ik-\zeta_{\pm} ,\frac{1}{2}-ik+\zeta_{\pm} ,1-ik
;\frac{1}{1+\textrm{e}^{-\frac{2\mu}{\theta_0}x}}\right),
\end{align}
where $\zeta_{\pm}=\frac{g\theta_0^2}{\mu}\pm\frac{1}{2}$, and $N_{\pm}^{k,\mathrm{L}}$ and $N_{\pm}^{k,\mathrm{R}}$ are the normalization factors for the continuum states.
We can easily check that the spinor constructed from solutions of the decoupled equations presented in Eq.\,(\ref{scatt}) or (\ref{scatt2}), with the same value of 
$k$, satisfies the original (coupled) Dirac Eq.\,(\ref{eom}) if we set 
$N^{k,\mathrm{L}}_-/N^{k,\mathrm{L}}_+=
-N^{k,\mathrm{R}}_-/N^{k,\mathrm{R}}_+
=(i\frac{\mu}{\theta_0}k+g\theta_0)
/\sqrt{\frac{\mu^2}{\theta^2_0}k^2+g^2\theta_0^2}$.
The asymptotic behavior of these wave functions at spatial infinities is as follows
\begin{equation}\label{asymp}
\psi^{(\pm)}_{k,\mathrm{L}}(x)=\begin{cases}
 N_{\pm}^{k,\mathrm{L}} \Gamma(1-ik)\left[\frac{\Gamma(-ik)\mathrm{e}^{ik\frac{\mu}{\theta_0}x}}
{\Gamma\left(\frac{1}{2}-ik +\zeta_{\pm}\right)\Gamma\left(\frac{1}{2}-ik-\zeta_{\pm}\right)}
 +\frac{\Gamma(ik)\mathrm{e}^{-ik\frac{\mu}{\theta_0}x}}{\Gamma\left(\frac{1}{2}+\zeta_{\pm}\right)
\Gamma\left(\frac{1}{2}-\zeta_{\pm}\right)}\right] ,&\mathrm{as}~x\rightarrow-\infty,\vspace{.3cm}\\
 N_{\pm}^{k,\mathrm{L}} \mathrm{e}^{ik\frac{\mu}{\theta_0}x},&\mathrm{as}~x\rightarrow+\infty,
\end{cases}
\end{equation}
\begin{equation}\label{asymp2}
\psi^{(\pm)}_{k,\mathrm{R}}(x)=\begin{cases}
 N_{\pm}^{k,\mathrm{R}}\mathrm{e}^{-ik\frac{\mu}{\theta_0}x}  ,&\mathrm{as}~x\rightarrow-\infty,\vspace{.3cm}\\
 N_{\pm}^{k,\mathrm{R}} \Gamma(1-ik)\left[\frac{\Gamma(ik)\mathrm{e}^{ik\frac{\mu}{\theta_0}x}}
{\Gamma\left(\frac{1}{2}+\zeta_{\pm} \right)
\Gamma\left(\frac{1}{2}-\zeta_{\pm}\right)}
+\frac{\Gamma(-ik)\mathrm{e}^{-ik\frac{\mu}{\theta_0}x}}
{\Gamma\left(\frac{1}{2}-ik +\zeta_{\pm}\right)
\Gamma\left(\frac{1}{2}-ik-\zeta_{\pm}\right)}\right],&\mathrm{as}~x\rightarrow+\infty.
\end{cases}
\end{equation}
This asymptotic behavior of $\psi^{(\pm)}_{k,\mathrm{L}}(x)$ (Eq.\,(\ref{scatt})), 
shown in Eq.\,(\ref{asymp}) represents an incident wave for $x\rightarrow -\infty$ moving to the right 
($\mathrm{e}^{ik\frac{\mu}{\theta_0}x}$), a reflected wave for $x\rightarrow -\infty$ moving back to the left ($\mathrm{e}^{-ik\frac{\mu}{\theta_0}x}$) 
and a transmitted wave for $x\rightarrow +\infty$ moving to the right ($\mathrm{e}^{ik\frac{\mu}{\theta_0}x}$).
As it is apparent from Eq.\,(\ref{asymp2}), the solution $\psi^{(\pm)}_{k,\mathrm{R}}(x)$ given in Eq.\,(\ref{scatt2}) represents the opposite scattering process.
By the use of these asymptotic behaviors, we can easily obtain the scattering phase shift for this model. 
Having obtained the phase shifts, we can use the strong form of the Levinson theorem for a double 
check on our calculations and the weak form of the Levinson theorem as a counter for the bound states.
In particular, we concentrate on the ever 
important zero mode.
\subsection{Phase shift and the Levinson theorem}
In this section we first find the phase shift for the scattering process embedded in Eqs.\,(\ref{scatt}-\ref{asymp2}).
To this end, we divide the coefficient of the transmitted wave by the coefficient of the incident wave and obtain 
the scattering matrix element which is related to the phase shift by the relation $S(k)=\mathrm{e}^{i\delta(k)}$.
Using Eq.\,(\ref{asymp}) or (\ref{asymp2}), the scattering matrix element of the fermion is as follows
\begin{align}\label{scatelement}
S_{\pm}(k)=\frac{\Gamma\left(\frac{1}{2}+\zeta_{\pm}-ik\right)\Gamma\left(\frac{1}{2}-\zeta_{\pm}-ik\right)}{\Gamma(-ik)\Gamma(1-ik)}=\mathrm{e}^{i\delta_{\pm}(k)},
\end{align}
where as before $\pm$ signs refer to the upper and lower components of the continuum wave function of the fermion.
As it can be seen, the $S$ matrix is ambiguous and the upper and lower components of the fermion wave function have different phase shifts.
We define the phase shift of the state to be  the average of the phase shifts of the two components 
of the fermion wave function \cite{levinsondr}. 
We shall show that this phase shift has all of the expected properties.
Also, notice that the $S$ matrix of our model and consequently the phase shifts of the fermion wave function 
are independent of the sign of the fermion energy, due to the symmetries of the system.
Therefore, the phase shifts of the states in the Dirac sea and sky with the same value of $|E|$ are the same. 
\par
\begin{center}
\begin{figure}[th] \hspace{1.cm}\includegraphics[width=13.cm]{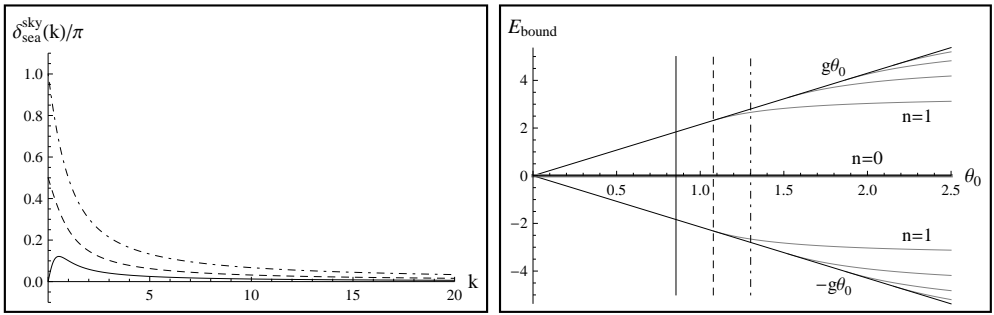}\caption{\label{phaseshift} \small
The left graph: 
The graphical representation of $\delta^{\mathrm{sky}}_{\mathrm{sea}}(k)/\pi$ as a function of $k$. 
In these graphs $g=2.15$ and $\mu=2.5$, and the values of $\theta_0$ for solid, 
dashed and dotdashed lines are $0.272 \pi$, $0.343242\pi$ and $0.414\pi$, respectively.
The right graph: The bound states of the fermion as a function of $\theta_0$. 
The three values of $\theta_0$ are shown by the solid, dashed and dotdashed lines.}
\end{figure}
\end{center}

Let us plot some examples of the phase shift of the fermion and explore them in connection with the Levinson theorem.
We concentrate on the case with the parameters $\mu=2.5$ and $g=2.15$ as the value of $\theta_0$ is increased from zero.
It is important to investigate the change in the values of the phase shifts at the boundaries of the continua, 
i.e.\;$\delta^{\textrm{sky}}(0)$ and $\delta^{\textrm{sea}}(0)$, when bound states appear.
In the left graph of Fig.\,\ref{phaseshift} we show the phase shifts as a function of $k$ for $\theta_0=\{0.272,0.343242,0.414\} \pi$. 
We draw the phase shifts for these three values of $\theta_0$ with solid, dashed and dotdashed lines, respectively.
In the right graph of this figure we show the fermion bound states as a function of $\theta_0$ and indicate the same 
three values of $\theta_0$ with solid, dashed and dotdashed lines.
As it can be seen, there are two threshold bound states (with $n=1$) at $\theta_0=0.343242\pi$.
At $\theta_0=0.272\pi$ there is only the zero-energy bound state and at $\theta_0=0.414\pi$ 
there are two additional bound states.
\par
We first utilize the weak form of the Levinson theorem for counting the exact number of the zero-energy bound states.
The weak form of this theorem for the Dirac equation can be written as follows (see for example \cite{levinsondr})
\begin{equation}\label{levweak}
\Delta \delta\equiv[\delta^{\mathrm{sky}}(0)-\delta^{\mathrm{sky}}(\infty)]+\left[\delta^{\mathrm{sea}}(0)-\delta^{\mathrm{sea}}(\infty)\right]
=\Delta\delta^{\textrm{sky}}+\Delta\delta^{\textrm{sea}}
=\left(N+\frac{N_\mathrm{t}}{2}-\frac{N_\mathrm{t}^0}{2}\right)\pi.
\end{equation}
Here $N$ is the total number of bound states, $N_\textrm{t}$ is the total number of threshold bound states at the given strength of the potential 
and $N_\textrm{t}^0$ is the total number of  threshold bound states at the zero strength of the potential.
In fact the last term takes into account the two threshold half bound states which exist for a free Dirac field in one spatial dimension.
Therefore, in this model $N_\textrm{t}^0=2$.
Now, we check this theorem for the sample phase shifts drawn in Fig.\,\ref{phaseshift}.
As shown in the left graph of this figure, for the phase shift at $\theta_0=0.272 \pi$, depicted by the solid line, 
we have $\Delta \delta^{\mathrm{sea}}/\pi=\Delta \delta^{\mathrm{sky}}/\pi=0$.
From the right graph we can see that $N_\mathrm{t}=0$ at $\theta_0=0.272 \pi$, and since $N_\mathrm{t}^0=2$, 
the Levinson theorem predicts only one bound state which is in fact the nondegenerate zero-energy bound state.
By considering the phase shift at the other values of $\theta_0$, we conclude the same result.  
\par
Now, we briefly explain the strong form of the Levinson theorem which deals with the value of the phase shift at $k=0$ ($E=\pm M_\textrm{f}$) and $k \to\infty$ ($E\to\pm\infty$), separately \cite{levinsondr}.
For $k=0$ this theorem can be expressed in the following form
\begin{equation}\label{levstrong1}
\delta(0)=\left(N_{\mathrm{exit}}-N_{\mathrm{enter}}\right)\pi.
\end{equation}
That is the value of the phase shift at zero momentum for each continuum is equal to the number of the bound states that exit minus the number of the bound states that enter that continuum from $E=\pm M_\mathrm{f}$, as the strength of the potential is increased from zero to its final value.
We can easily see that the sample phase shifts shown in Fig.\,\ref{phaseshift} are consistent with the relation 
(\ref{levstrong1}).
For example, at $\theta_0=0.272\pi$ the value of the phase shifts at zero momentum ($E=\pm M_\mathrm{f}$) are $0$.
On the other hand, from the right graph of Fig.\,\ref{phaseshift} we see the total number of 
bound states that have exited each of the continua from the line $E=\pm M_\mathrm{f}$ is $0$.
We conclude that the ever present zero-energy bound state has been formed from the union of the two 
threshold (half) bound states present in the free case, i.e.\;at $\theta_0=0$.
At $\theta_0=0.343242\pi$ two threshold bound states have formed at $E=\pm M_{\textrm{f}}$ and the corresponding 
phase shifts are $\delta^{\textrm{sky}}_{\textrm{sea}}(0)=\frac{\pi}{2}$.
For $\theta_0=0.404\pi$ these states have completely separated from the continua and become full bound states.
Therefore, the corresponding phase shifts attain the value $\pi$. 
This sort of consistencies
can be easily seen for other phase shifts.
\par
The strong form of the Levinson theorem for $k \to\infty$ can be written as follows
\begin{equation}\label{levstrong2}
\delta(\infty)=\left(N_{\mathrm{enter}}-N_{\mathrm{exit}}\right)\pi.
\end{equation}
This means that the value of the phase shift for $E\to\pm\infty$ is equal to the total number of the bound states 
that enter minus the number of the bound states that exit that continuum from $E=\pm\infty$, as the strength 
of the potential is increased from zero to its final value.
For the generic cases where the presence of a background field does not cause the Hamiltonian to lose its 
hermiticity the spectrum remains complete, and hence 
$\delta^{\textrm{sky}}(\infty)+\delta^{\textrm{sea}}(\infty)=0$.
Since in this model the charge and particle conjugation symmetrirs
imply that $\delta^{\textrm{sky}}(k)=\delta^{\textrm{sea}}(k)$ for all $k$, we expect 
$\delta^{\textrm{sky}}(\infty)=\delta^{\textrm{sea}}(\infty)=0$, and this is precisely the results shown in 
Fig.\,\ref{phaseshift}. 
Now we emphasis several important properties of the spectrum of the model at this point.
As is evident from the right graph of Fig.\,\ref{phaseshift}, 
there are no level crossing, the spectrum is symmetric about $E=0$, the dynamical mass generated is 
$M_\textrm{f}=g\theta_0$, and the zero mode is always present.
\section{Conclusion}
In this paper we thoroughly investigate the ($1+1$)-dimensional model considered by Jackiw and Rebbi, whence
they have introduced the possibility of the fractional fermion number for the ground state.
In this model a Fermi field is coupled to a prescribed pseudoscalar field in the form 
of the kink characterized by two parameters $\theta_0$ 
(the value of the kink at spatial infinity) and $\mu$ 
(the slope of the kink at $x=0$).
We solve the equations of motion of this system analytically and exactly, and find the 
bound state wave functions and energies as well as the continuum states for arbitrary 
choice of the parameters $\theta_0$ and $\mu$.
Then, we plot the bound state energies of the fermion as a function of the parameter 
$\theta_0$.
Having this complete set of solutions, we can consider a process in which the background field 
evolves from zero to its final form.
We can then observe the changes in the spectrum of the fermion during this process.
We find that the interaction induces a mass for the Fermi field $M_{\textrm{f}}=g\theta_0$.
That is although the free theory is massless, a band gap appears as $\theta_0$ increases.
Also, the bound states start appearing with $N_{\mathrm{b}}=2\left[\frac{g\theta_0^2}{\mu}\right]_{\mathrm{F}}+1$.
The invariance of the system under the charge and particle conjugation symmetry 
is obvious in the graph of the bound energies, since the energy levels are totally 
symmetric with respect to the line of $E=0$.
We focus our attention especially on the zero-energy fermionic mode and see that 
this mode is always present for every choice of the parameters.
In particular, we expose the origin of the ever present zero mode: 
In the free Dirac case, i.e.\;when $\theta_0=0$, there is no mass gap and two threshold zero-energy bound states seperate the continua.
As $\theta_0$ increases, a mass gap appears and the two threshold bound states merge to form the zero-energy bound state.
We also find the continuum scattering states of the fermion analytically and then 
calculate the phase shift of these wave functions.
Using the weak form of the Levinson theorem, we conclude that the self charge conjugate zero-energy 
fermionic mode is a nondegenerate mode as Jackiw and Rebbi stated.  
 
\section*{Acknowledgement} We would like to thank the research office
of the Shahid Beheshti University for financial support.



\begin{thebibliography}{9}
\bibitem{jackiw-rebbi}
R. Jackiw and C. Rebbi, Phys. Rev. \textbf{D 13}, 3398 (1976).

\bibitem{jackiw}
R. Jackiw, Rev. Mod. Phys. \textbf{49}, 681 (1977).

\bibitem{goldstone}
J. Goldstone and F. Wilczek, Phys. Rev. Lett. \textbf{47}, 986 (1981).

\bibitem{mackenzie1}
R. MacKenzie and F. Wilczek, Phys. Rev. \textbf{D 30}, 2194 (1984).

\bibitem{mackenzie2}
R. MacKenzie and F. Wilczek, Phys. Rev. \textbf{D 30}, 2260 (1984).

\bibitem{dr}
S. S. Gousheh and R. L\'{o}pez-Mobilia, Nucl. Phys. \textbf{B 428}, 189 (1994).

\bibitem{leila}
L. Shahkarami and S. S. Gousheh, JHEP \textbf{06}, 116 (2011).

\bibitem{dehghan}
Z. Dehghan and S. S. Gousheh, Int. J. Mod. Phys. \textbf{A 27}, 1250093 (2012).

\bibitem{cosm1}
E. R. Bezerra de Mello and A. A. Saharian, Phys. Rev. \textbf{D 75}, 065019 (2007).

\bibitem{cosm2}
E. R. Bezerra de Mello, V. B. Bezerra, A. A. Saharian and A. S. Tarloyan, Phys. Rev. \textbf{D 78}, 105007 (2008).

\bibitem{cosm3}
E. R. Bezerra de Mello and A. A. Saharian, Phys. Rev. \textbf{D 78}, 045021 (2008).

\bibitem{cosm4}
E. R. Bezerra de Mello and A. A. Saharian, J. Phys. \textbf{A 45}, 115002 (2012).

\bibitem{cosm5}
E. R. Bezerra de Mello, A. A. Saharian and S. V. Abajyan, Class. Quant. Grav. \textbf{30}, 015002 (2013).

\bibitem{cond1}
W. P. Su, J. R. Schrieffer and A. J. Heeger, Phys. Rev. Lett. \textbf{42}, 1698 (1979).

\bibitem{cond2}
W. P. Su and J. R. Schrieffer, Phys. Rev. Lett. \textbf{46}, 738 (1981).

\bibitem{cond3}
A. Niemi and G. Semenoff, Phys. Rep. \textbf{135}, 99 (1986). 

\bibitem{cond4}
J. Ruostekoski, J. Javanainen and G. V. Dunne, Phys. Rev. \textbf{A 77}, 013603 (2008).

\bibitem{poly1}
M. Rice and E. Mele, Phys. Rev. Lett. \textbf{49}, 1455 (1982).

\bibitem{poly2}
R. Jackiw and G. Semenoff, Phys. Rev. Lett. \textbf{50}, 439 (1983).

\bibitem{poly3}
A. J. Heeger, S. Kivelson, J. R. Schrieffer and W.-P. Su, Rev. Mod. Phys. \textbf{60}, 781 (1988).

\bibitem{atom1}
A. R. Neghabian, Phys. Rev. \textbf{A 27}, 2311 (1983).

\bibitem{atom2}
Y. Gu, Phys. Rev. \textbf{A 66}, 032116 (2002).

\bibitem{atom3}
A. I. Milstein, I. S. Terekho, U. D. Jentschura and C. H. Keitel, Phys. Rev. \textbf{A 72}, 052104 (2005).

\bibitem{dashen}
R. F. Dashen, B. Hasslacher and A. Neveu, Phys. Rev. \textbf{D 10}, 4130 (1974).

\bibitem{rajaraman}
R. Rajaraman, \textit{Solitons and Instantons: An Introduction to Solitons and Instantons in Quantum Field Theory} 
(North-Holland, Amsterdam, 1982).

\bibitem{shape1}
L. E. Gendenshtein, JETP Lett. \textbf{38}, 356 (1983).

\bibitem{shape2}
F. Cooper and B. Freedman, Ann. Phys. (N.Y.) \textbf{146}, 262 (1983).

\bibitem{shape3}
F. Cooper, A. Khare and U. Sukhatme, Phys. Rep. \textbf{251}, 267 (1995).

\bibitem{shape4}
G. Junker, \textit{Supersymmetric Methods in Quantum and Statistical Physics} (Springer, Berlin, 1996).

\bibitem{shape5}
J. Sadeghi and A. Mohammadi, Eur. Phys. J. \textbf{C 49}, 859 (2007).

\bibitem{shape6}
A. A. Andrianov and M. V. Loffe, J. Phys. \textbf{A 45}, 503001 (2012).

\bibitem{shape7}
A. Alonso-lzquierdo, G. M. Guilarte and M. S. Plyushchay, Ann. Phys. (Amsterdam) \textbf{331}, 269 (2013).

\bibitem{levinson}
N. Levinson, Kgl. Danske Videnskab. Selskab. Mat.-fys. Medd. \textbf{25}, 1 (1949).

\bibitem{levinsondr}
S. S. Gousheh, Phys. Rev. \textbf{A 65}, 032719 (2002).

\bibitem{arxiv}
S. Kutnii, arXiv:hep-th/1107.1889v1.

\bibitem{morse}
P. M. Morse and H. Feshbach, \textit{Methods of Theoretical Physics, Vol.\;II} (McGraw-Hill, 1953).

\bibitem{landau}
L. D. Landau and E. M. Lifshitz, \textit{Quantum Mechanics: Non-relativistic Theory} (Pergamon Press, 1989).

\end{thebibliography}
 \end{document}